\documentclass[sigconf]{acmart}
\AtBeginDocument{%
  \providecommand\BibTeX{{%
    \normalfont B\kern-0.5em{\scshape i\kern-0.25em b}\kern-0.8em\TeX}}}

\usepackage{booktabs}
\usepackage{graphicx, wrapfig}

\def\authnote{1} 
\usepackage{xcolor}
\usepackage{natbib}
\usepackage{tabularx}
\newcolumntype{Y}{>{\arraybackslash}X}

\renewcommand{\paragraph}[1]{\vspace*{6pt}\noindent\textbf{#1.}\;}

\newcommand{\secref}[1]{Section~\ref{#1}}

\newcommand{\fixme}[1]{\ifnum\authnote=1{\textcolor{red}{[FIXME: #1]}}\fi}
\newcommand{\better}[1]{\ifnum\authnote=1{\textcolor{violet}{[BetterWord: #1]}}\fi}
\newcommand{\todo}[1]{\ifnum\authnote=1{\textcolor{red}{\textbf{[TODO: #1]}}}\fi}

\newcounter{mynote}[section]

\newcommand{\thenote}{\thesection.\arabic{mynote}}

\newlength{\saveparindent}
\newlength{\saveparskip}
\newcounter{ctr}

\newcommand{\enote}[1]{\ifnum\authnote=1\refstepcounter{mynote}{\textbf{\textcolor{blue}{$\ll$ET~\thenote: {\sf #1}$\gg$}}}\fi}

\newcommand{\hnote}[1]{\ifnum\authnote=1\refstepcounter{mynote}{\textbf{\textcolor{red}{$\ll$HS~\thenote: {\sf #1}$\gg$}}}\fi}

\newcommand{\tnote}[1]{\ifnum\authnote=1\refstepcounter{mynote}{\textbf{\textcolor{orange}{$\ll$MY~\thenote: {\sf #1}$\gg$}}}\fi}

\newcommand{\rev}[1]{\ifnum\revview=1{\color{blue}{#1}}\else{#1}\fi}

\copyrightyear{2024}
\acmYear{2024}
\setcopyright{rightsretained}
\acmConference[FAccT '24]{The 2024 ACM Conference on Fairness, Accountability, and Transparency}{June 3--6, 2024}{Rio de Janeiro, Brazil}
\acmBooktitle{The 2024 ACM Conference on Fairness, Accountability, and Transparency (FAccT '24), June 3--6, 2024, Rio de Janeiro, Brazil}\acmDOI{10.1145/3630106.3658992}
\acmISBN{979-8-4007-0450-5/24/06}

\begin{document}

\title{Participation in the Age of Foundation Models}

\author{Harini Suresh}
\authornote{These authors contributed equally to this research.}
\orcid{0000-0002-9769-4947}
\affiliation{%
  \institution{Cornell Tech}
  \country{USA}
}

\author{Emily Tseng}
\authornotemark[1]
\affiliation{%
  \institution{Cornell University}
  \country{USA}
}

\author{Meg Young}
\authornotemark[1]
\affiliation{%
  \institution{Data \& Society Research Institute}
  \country{USA}
}

\author{Mary L. Gray}
\affiliation{%
  \institution{Microsoft Research}
  \country{USA}
}

\author{Emma Pierson}
\affiliation{%
  \institution{Cornell Tech}
  \country{USA}
}

\author{Karen Levy}
\affiliation{%
  \institution{Cornell University}
  \country{USA}
}

\renewcommand{\shortauthors}{Suresh, Tseng, Young et al.}

\begin{abstract}
Growing interest and investment in the capabilities of \textit{foundation models} has positioned such systems to impact a wide array of services, from banking to healthcare.  Alongside these opportunities is the risk that these systems reify existing power imbalances and cause disproportionate harm to historically marginalized groups. The larger scale and domain-agnostic manner in which these models operate further heightens the stakes: any errors or harms are liable to reoccur across use cases.  In AI \& ML more broadly, \textit{participatory} approaches hold promise to lend agency and decision-making power to marginalized stakeholders, leading to systems that better benefit justice through equitable and distributed governance. 
But existing approaches in participatory AI/ML are typically grounded in a specific application and set of relevant stakeholders, and it is not straightforward how to apply these lessons to the context of foundation models.
Our paper aims to fill this gap.

First, we examine existing attempts at incorporating participation into foundation models. We highlight the tension between participation and scale, demonstrating that it is intractable for impacted communities to meaningfully shape a foundation model that is intended to be universally applicable. In response, we develop a blueprint for participatory foundation models that identifies more local, application-oriented opportunities for meaningful participation.  In addition to the ``foundation'' layer, our framework proposes the ``subfloor'' layer, in which stakeholders develop shared technical infrastructure, norms and governance for a grounded domain such as clinical care, journalism, or finance, and the ``surface'' (or application) layer, in which affected communities shape the use of a foundation model for a specific downstream task.
The intermediate ``subfloor'' layer scopes the range of potential harms to consider, and affords communities more concrete avenues for deliberation and intervention. At the same time, it avoids duplicative effort by scaling input across relevant use cases.
Through three case studies in clinical care, financial services, and journalism, we illustrate how this multi-layer model can create more meaningful opportunities for participation than solely intervening at the foundation layer.

\end{abstract}




\maketitle

\section{Introduction}
Recent years have seen a notable rise in interest and investment into \textit{foundation models} \cite{bommasani2021opportunities}, exemplified by systems such as GPT-4 or CLIP.  Foundation models are unique in their generalizability, with the ability to adapt to a range of tasks not explicitly introduced during training. 
While many of the methods underpinning foundation models are not new (e.g., pre-training via self-supervised learning on unlabeled data), they are now being developed and deployed at an unprecedented scope and scale. 
These systems have spurred broad interest across industries including medicine \cite{epic2023generative,clusmann2023future}, software engineering \cite{wang2023software}, and education \cite{extance2023chatgpt}. 
Alongside these opportunities, new and heightened risks have emerged, including environmental \cite{luccioni2023power, strubell2019energy} and economic impacts \cite{brynjolfsson2023generative}, extractive data labor \cite{Perrigo2023}, legal concerns \cite{lee2023talkin}, data inscrutability \cite{bender2021dangers}, and the homogenization of discriminatory behavior across applications \cite{bommasani2021opportunities}. 

Of particular concern is that these risks and opportunities will disproportionately impact different groups---further advantaging those who already benefit from existing power structures, while historically marginalized communities bear the brunt of the resulting harms \cite{bender2021dangers}.  
This concern---where by default, technological systems operate within and reflect back systems of structural oppression and power---has long been discussed in the context of technology and ML systems more broadly \cite{d2023data,benjamin2023race,noble2018algorithms}. 
To mitigate these power imbalances, there have been increasing calls for more \textit{participation} in ML, i.e., for a broader range of people and communities to be involved in shaping whether and how systems are built and deployed \cite{kulynych2020participatory,delgado2023participatory,OpenAI_2023_behave,wolf2018changing}. 
In recent years, participatory ML efforts have gained traction, spanning applications such as machine translation for low-resourced languages \cite{hao2022new,cofey2021maori,nekoto-etal-2020-participatory}, matching algorithms for food donation services \cite{lee2019webuildai}, and news classification systems to support activists monitoring gender-related violence \cite{suresh2022towards}.  
Alongside these endeavors, scholars have also described how participatory efforts can fall short of meaningfully shifting power to the marginalized, such as through \textit{participation-washing} or co-optation by powerful interests \cite{sloane2022participation,birhane2022power,delgado2023participatory,corbett2023power}. 

Given the scale at which foundation models are already affecting society and capturing imagination, it is critical for technologists to actively mitigate the power imbalances they produce, as well as the disproportionate harms. But \textbf{how can the benefits of participatory ML be realized within the unique affordances of foundation models?} 
In a classic ML setup, the downstream task and likely users are known upfront, and participatory ML approaches might look like community control over problem formulation, data collection and storage, or the design of evaluations that best reflect real-world use. 
However, this is much more challenging in the foundation model paradigm, where downstream use cases and stakeholders are disconnected from the model---both conceptually (the model is intended to perform well on an unbounded set of potential use cases) and practically (development primarily happens within large tech companies that are not accountable to specific communities).  If valuing local expertise and context are core to participatory methods, can foundation models be meaningfully participatory?

In this paper, we investigate the proposition of participatory foundation models by first examining existing efforts that attempt to incorporate stakeholder input into foundation model development (\secref{sec:ceiling-overall}). By analyzing them through the lens of participatory scholarship, we find consistent limitations in the ability of these mechanisms to meaningfully shift power, highlighting the tension between participation and scale. Based on these findings, we conceptualize the \textit{participatory ceiling}: an inherent limit on the ability for impacted communities to meaningfully shape a foundation model that is intended to be almost universally applicable.  

Then, we develop a blueprint for participation in foundation models that identifies more local, application-oriented opportunities to lend stakeholders meaningful agency and decision-making power (\secref{sec:blueprint}).  The framework extends the carpentry metaphor of a \textit{foundation}---an unfinished base that can support many different kinds of structures.  Built on top of the foundation, a \textit{subfloor} layer provides a level and structurally-sound base to support the top-level \textit{surface} layer.  In our framework, the subfloor layer
encompasses technical infrastructure, norms, and/or governance for a grounded domain (e.g., reproductive health).  It helps scope the range of potential uses to consider, lend clarity to who should be involved, and ground considerations of harm and equity in the sociohistorical context of a domain.  
The surface layer, then, builds on the subfloor, and encompasses specific downstream use cases (e.g., a chatbot to help provide patients with accurate information on fertility concerns).  The surface layer provides opportunities for task- and locale-specific participatory engagement---while also benefiting from the domain-specific infrastructure, norms and governance developed at the subfloor layer. We walk through three case studies in clinical care, financial services, and journalism, to illustrate how this multi-layer framework can support public power and decision-making over ML in the foundation model paradigm.






\section{Related Work}
\label{sec:relwork}

\subsection{Participatory machine learning}
\label{sec:relwork-participation}
Participatory traditions have a long history in and outside of technology design  \cite{minkler2011community,mcintyre2007participatory,arnstein1969ladder,fung2006varieties,muller1993participatory,greenwood2006introduction}. Specific motivations for participation vary, and include redistributing decision-making power to those who have less \cite{muller1993participatory}, learning from participants' expertise or preferences \cite{feffer2023preference,patnaik1999needfinding}, or meeting epistemic goals such as procedural fairness \cite{greenberg1983procedural}.
In this paper, we focus on the promise of participatory approaches to shift power to those with less, following a range of participatory scholarship centering issues of power, agency and accountability \cite{arnstein1969ladder,gregory2003scandinavian,costanza2020design,palacin2020design,cooke2001participation,sloane2022participation}.

In AI and ML, 
recent work has applied participatory methods at various points throughout the ML lifecycle, including problem formulation, data collection and annotation, model development, evaluation, and governance \cite{lee2019webuildai, nekoto-etal-2020-participatory, suresh2022towards, katell2020toward, Queerinai2023,hao2022new,cofey2021maori}. 
Alongside these examples, research has also pointed out pitfalls of ``participation-washing.''  \citet{sloane2022participation} distinguish \textit{participation as work} (e.g., user data used to train models without consent or compensation) and \textit{participation as consultation} (e.g., one-off focus groups to elicit user preferences) from \textit{participation as justice} (sustained and mutually beneficial relationships with communities, who co-determine if and how a model should be built).  This gradation aligns with historical perspectives on the co-optation of participation, such as Arnstein's ``Ladder of Citizen Participation,'' which describes the degree to which political and economic processes redistribute power to citizens \cite{arnstein1969ladder}.  

Several recent papers have used additional frameworks and heuristics to understand the range of participatory ML work.
\citet{corbett2023power} use Arnstein's ladder to compare and contrast the extent to which different participatory approaches redistribute power, highlighting eight case studies of each rung on the ladder.  \citet{delgado2023participatory} synthesize the ladder with other participatory scholarship to develop the ``Parameters of Participation,'' a framework they then use to analyze the goals and scope of 80 research papers in participatory AI. \citet{birhane2022power} describe further axes of interest (reflexivity, empowerment, reciprocity, duration), using these to analyze three case studies \cite{birhane2022power}. 
Focusing on commercial AI labs, \citet{groves2023going} interview industry practitioners, illustrating barriers that arise when attempting to use participatory approaches in industrial practice (e.g., a lack of resources or misaligned incentives). 

Thus far, the participatory ML literature has focused on small-scale, application-focused case studies.
This is expected,
as scholarship on participatory approaches encourages context-specificity.
However, it leaves us with a gap between how participatory ML is being conceptualized, and the increasing prevalence of foundation models that operate on much larger scales. 
To address this gap, our paper builds on the analysis and heuristics developed in these prior works to characterize the landscape of participatory efforts in the foundation model ecosystem, and present a framework for more participatory alternatives.

\subsection{Foundation models}
\label{sec:relwork-foundation}
Coined in 2021, the term \textit{foundation model} \cite{bommasani2021opportunities} refers to a paradigm of ML in which a base model (the \textit{foundation}) is trained via self-supervised learning to compactly represent the statistical distribution of a vast dataset (\textit{pretraining}). 
Pretrained representations (also called \textit{embeddings}) can then be adapted to downstream tasks (\textit{fine-tuning} for specific \textit{applications}).
Pretraining-finetuning predates \citet{bommasani2021opportunities}; we adopt the foundation model terminology here to expand on its metaphor, and speak to the current era of large and centralized models, including both closed-source (e.g., OpenAI's GPT) and open-source (e.g., DBRX, Llama) variants.

A foundation model approach offers application developers tremendous advantages---and offers foundation model developers tremendous power. 
Instead of training their models with as large of a dataset as they can find, store, and compute, an application developer may take an available foundation model (e.g., one trained on vast amounts of Internet text to represent the English language) and adapt it for their task, thus inheriting the knowledge representations learned during pretraining.
Achieving the largest and most performant foundation model has thus become an arms race within the machine learning community.
Institutions like tech companies or well-resourced universities look to amass vast stores of data---e.g., LLMs require datasets in the range of billions of words---that is typically only available at the scale of Internet-wide scrapes \cite{bender2021dangers}. 
Training and deploying these models similarly requires massive amounts of compute, which comes with growing environmental costs \cite{strubell2019energy, luccioni2023power}: foundation models today can feature trillions of parameters.
Nevertheless, the race accelerates. 
Since the initial set of foundation models cited in \citet{bommasani2021opportunities}, including BERT, GPT, and CLIP, foundation models have seen rapid uptake in the tech industry, and a new wave of startups has emerged, all aiming to commercialize the best foundation models, and capture the widest share of downstream applications.

The foundation model paradigm poses fundamental challenges for the participatory ML approaches built for the task-specific era. 
Foundation models are inherently context-less; they are trained to simply represent a dataset, and provide baseline performance for an infinite horizon of future tasks. Moreover, due to their universal aim and the resources needed to develop them, foundation models are primarily built and controlled by large tech companies that do not have a particular investment in or accountability to any given community or domain.  These departures from the task-specific paradigm pose challenges for participation at each step of ML lifecycle --- from problem definition to data collection to deployment.  
For example, dataset auditing may have previously been informed by a domain-specific understanding of the data generation and curation processes.  For a foundation model where the data is intended to inform a universally-applicable knowledge representation, specifying and finding dataset harms becomes a more nebulous problem.  And this is before considering the lack of transparency into those datasets by foundation model providers \cite{bommasani2023foundation}.  Indeed, numerous harms of foundation model datasets have emerged, including LLMs that memorize and/or leak private information \cite{brown2022does, nasr2023scalable}, or image generation models that rely on datasets that contain hateful or even illegal content (such as child sexual abuse material) \cite{birhane2023into}.


There are clear challenges to building meaningful participation into the foundation model paradigm. While fine-tuning may address some of these challenges, a fine-tuned model inherits the limitations of a foundation model, including its biases, and it is not yet clear how a community would participate in ensuring a fine-tuned model was fit-for-purpose.
Our work interrogates the possibility of participatory approaches to foundation model development, towards mitigating harm in this new scientific and industrial paradigm.

\section{The Participatory Ceiling in Foundation Models}
\label{sec:ceiling-overall}

We begin by reviewing proposed mechanisms for participation in foundation model development.
We analyze these attempts through the lens of the Parameters of Participation, a conceptual framework developed by \citet{delgado2023participatory} to characterize 
participatory initiatives in AI and ML.  
Based on this analysis, we argue there is a \textit{participatory ceiling} that limits the extent to which participatory approaches can meaningfully redistribute decision-making power when directly intervening on a foundation model. 

\subsection{The Parameters of Participation}
We use \citet{delgado2023participatory}'s 
Parameters of Participation as a conceptual framework because it is focused on participatory AI efforts in particular, in contrast to more general theories or frameworks of participation.  
The framework 
articulates key \textit{dimensions} along which participatory approaches differ.  
The dimensions, also framed as questions, include the \textbf{goal} (why is participation needed?), the \textbf{stakes} (what is on the table?), the \textbf{scope} (who is involved?), and the \textbf{form} (what form does participation take?).  The answers to these questions are structured along a spectrum of four \textit{modes} of participation, which span \textbf{consultation} (e.g., eliciting user preferences), \textbf{inclusion} (e.g., deliberation around specific design choices), \textbf{collaboration} (e.g., co-creation of design possibilities), and \textbf{ownership} (e.g., stakeholders shape the entire design process). 
This spectrum of modes reflects a long tradition in participatory scholarship that encourages engagements that cede a greater degree of decision-making power to those most directly affected by the outcome (i.e., more ``meaningful participation'') \cite{arnstein1969ladder,fung2006varieties,iap2}.

\subsection{How participatory are existing participatory foundation model efforts?}
We used purposive sampling \cite{patton2014qualitative} to review academic and gray literature on approaches to using participatory methods or promoting public input for foundation models. Relevant examples were identified from websites of foundation model providers (e.g., OpenAI blog posts), AI/ML conference proceedings, and arXiv. 
We inductively clustered examples into several broad categories, including RLHF, methods that develop rulesets, guidelines or policies, and red teaming. We then applied \citeauthor{delgado2023participatory}'s Parameters of Participation as an analytical lens to characterize these approaches by the degree to which they afford meaningful participation, or how they might deepen along that axis \cite{delgado2023participatory}. A summary of these findings is illustrated in Figure 1.



There are a few forms of human input we did not include in our analysis.  
First, while all development hinges on human decisions and norms, we consider participation to involve some kind of public external to the model development team, and excluded approaches that were limited to developer feedback.
In addition, we do not include the generic use of human-generated data as a mechanism for human input.  This data is certainly important for foundation models, which, like many ML models, rely on ``human infrastructures'' of annotators, data workers, and content created by people \cite{mateescu2019ai}.  We defer to existing literature that has showed how this mode of input is nominal and extractive \cite{miceli2020between, miceli2022data, mateescu2019ai, sloane2022participation}, and instead focus on approaches that intend to give participants more agency (e.g., red teaming) or that are unique to foundation models (e.g., RLHF). 


\begin{figure*}
    \centering
    \includegraphics[width=0.8\textwidth]{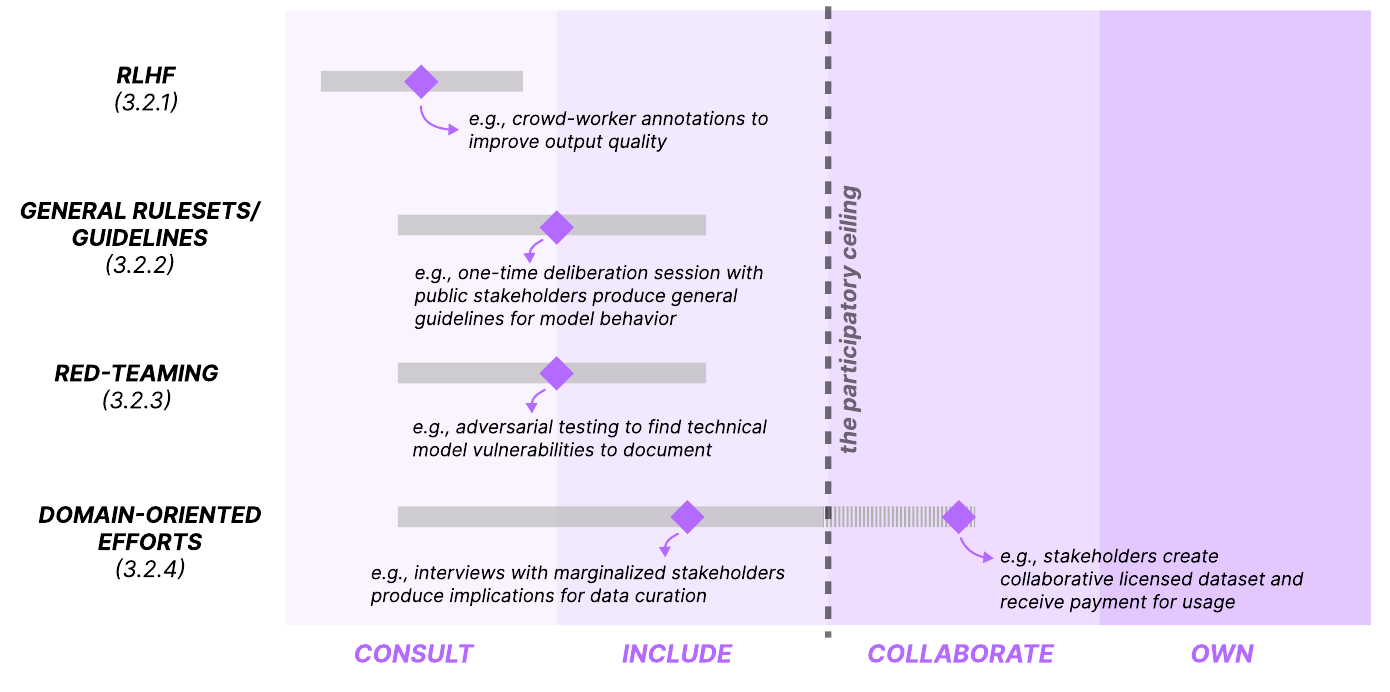}
    \caption{Different categories of participatory approaches, along with the modes of participation (as described in  \cite{delgado2023participatory}) they currently cover.  The bars are a qualitative depiction of the modes covered by each method -- e.g., dimensions for red teaming approaches we reviewed fell equally under consult and include; so the range spans both modes equally. The purple diamonds are exemplars of each category.  In the domain-oriented efforts category, we illustrate the possibility for a few approaches to pass the participatory ceiling with the dashed bar.  We describe these exploratory approaches in \secref{sec:ceiling}.}
    \label{fig:enter-label}
\end{figure*}

\subsubsection{Reinforcement learning with human feedback (RLHF)}
\label{sec:rlhf}
A primary mechanism for improving the quality of foundation model outputs is reinforcement learning with human feedback (RLHF) \cite{christiano2023deep, stiennon2022learning, ouyang2022training, ziegler2020finetuning, bai2022training, sharma2023understanding, askell2021general, kreutzer-etal-2021-offline, yang2023using, bakker2022fine, rafailov2023direct,kirk2023past}.  
As currently practiced, participants in RLHF are typically crowd-workers or contractors; most documentation does not provide further detail on recruiting strategies or demographics. 
The form of feedback is typically through online questionnaires where participants assign a comparative rating to two model outputs.  This feedback is used to train a ``preference model,'' which serves as a reward function in a subsequent reinforcement learning training procedure.
As it is most commonly instantiated, RLHF is limited to the \textit{consult} mode of participation, in all dimensions. Its primary goal is to improve the quality of the model according to desiderata set by the development team. The scope of participants is limited to those who can give a prescribed quantity of discrete feedback, rather than those who can give specific kinds of expertise. The form of feedback is a single lever, typically ranking pairs of outputs. And the stakes are limited to adjusting a pre-trained model.

\subsubsection{Rulesets and policies} 
\label{sec:rulesets}
A variety of approaches aim to synthesize principles, rules, or policies about foundation model behavior based on human input or deliberation.  
For example, Anthropic's Collective Constitutional AI (CCAI) uses a public polling process to determine a ruleset (also referred to as a ``constitution'') \cite{anthropic2023ccai}, which takes the form of a set of instructions for the model that reflects specific values or principles (e.g., ``choose the response that is most respectful'').  Participants (in this case, ``a representative sample of 1,000 U.S. adults across age, gender, income, and geography'') vote on candidate principles, or submit their own; those with high consensus make it into the ruleset.  Similarly, many of the projects funded by OpenAI's ``Democratic Inputs to AI'' grant program develop methodologies for collecting and consolidating public input to produce a set of representative beliefs or statements \cite{Eloundou_Lee_2024, konya2023democratic,theuns2023deliberation,krasodomski2023vtaiwan, mendoza2023making, wang2023inclusive, fish2023generative, shaotran2023aligned}.  Participants are typically identified and recruited by the research team, and range from representative samples of the public to more demographically- or geographically-focused groups.  The forms of participation in these approaches include voting or contributing free-form thoughts on online platforms \cite{konya2023democratic,krasodomski2023vtaiwan,fish2023generative,shaotran2023aligned,edelman2023democratic}, deliberation in online chat rooms \cite{wang2023inclusive}, and focus groups or interviews \cite{mendoza2023making,theuns2023deliberation,bergman2024stela}.  In many cases, foundation models are further used as a tool in producing the final output---e.g., LLMs facilitate online discussions \cite{mendoza2023making} or summarize participant input \cite{shaotran2023aligned}.  The outputs of these methods are used in different ways.  Sometimes, as in the case of CCAI, the resulting ruleset is used to steer the model via a reinforcement learning procedure;  in many other cases, there is not explicit guidance on how the resulting documentation or findings should be used.    

The goal of rulesets is typically to adjust the model to reflect stakeholder preferences and values.  
The scope of participants is determined by the project team according to a particular value (e.g., representativeness). Along these dimensions, synthesizing public input into rulesets or guidelines falls under the \textit{include} mode of participation. At the same time, the stakes are low. Participants can vote on constitutional principles, or contribute to discussions around model behavior. 
But how people's inputs end up impacting the model is neither guaranteed nor transparent, mediated by complicated RL processes or LLM-facilitated analyses.  
And the \textit{form} of participation---providing preferences or input during a discrete time window---reflects the fact that ultimately, participants have little say regarding the model's impact in the world: whether it is developed, what other data it is trained on, what it may be used for, or if and how it should be deployed.  Some methods aim to produce guidelines that touch on these broader questions, but to have impact, they require additional mechanisms of downstream control that are not currently in place.  In short, with respect to the stakes and form of participation, these approaches remain in the \textit{consult} mode of participation.

\subsubsection{Red teaming}
\label{sec:redteaming}
Red teaming, a practice developed in computer security, involves enlisting domain experts to adversarially test systems to uncover specific kinds of weaknesses.
Recent efforts at foundation model governance have explored how red teaming can be applied to AI: e.g., in September 2023, OpenAI announced a call for a red teaming network, made up of external domain experts, ``to help develop domain specific taxonomies of risk and evaluating possibly harmful capabilities in new systems'' \cite{OpenAI_2023}. Here, participants are recruited by the project team for their domain expertise in predefined areas (typically, along with technical expertise), consistent with the \textit{inclusion} mode of participation.  Red teaming programs have involved individuals interacting with a model, group discussions, and shared documentation of findings from weeks of use \cite{Dalle2022,gpt2023}. 
When red teaming involves deliberation and discussion with the project team, it becomes \textit{inclusion}, which improves upon methods like online surveys (which remain at the \textit{consultation} level), but does not yet achieve the \textit{collaboration} mode, which would involve more durable relationships and decision-making power.
Red teaming is also usually scoped to uncovering specific technical vulnerabilities. Therefore, the stakes are limited to \textit{consultation}. Findings from a red teaming engagement might inform adjustments to the model or its documentation on harms, but these decisions are still ultimately up to the development team.


\subsubsection{Domain-oriented efforts}
\label{sec:domainharms}
Other approaches---while similar in many ways to creating rulesets with public input (\secref{sec:rulesets}) or red teaming (\secref{sec:redteaming})---stem more from understanding risks that are relevant to a particular domain or group.  For example, researchers have engaged stakeholders in maternal health fields \cite{antoniak2023designing}, creative professions \cite{inie2023designing}, or disability \cite{gadiraju2023wouldn} and queer communities \cite{felkner2023winoqueer} about risks or opportunities for foundation models that are important to them.  Most commonly, findings result in high-level guidelines for usage or harm measurement.  For example, \citet{antoniak2023designing} propose guiding principles for using LLMs in maternal health applications.  In other cases, findings shape evaluative benchmarks \cite{felkner2023winoqueer}.  And in one exploratory example we encountered, participants (in this case, African artists) contributed to a collaborative licensable dataset with a payout structure for compensating creators \cite{eglash2023ubuntu}.

For the most part, as in Sections \ref{sec:rulesets} and \ref{sec:redteaming}, these efforts reflect the \textit{inclusion} mode of participation along each dimension.  But some aim to expand the stakes, raising issues related to data quality and acceptable use cases \cite{antoniak2023designing,gadiraju2023wouldn}, or designing interventions targeting data collection and curation \cite{eglash2023ubuntu}. These opportunities for more meaningful participation arise, in part, through the focus on a grounded domain.  That said, these approaches are exploratory, and it remains challenging for domain-specific stakeholders to influence the foundation model with any decision-making power.  We expand on this challenge in \secref{sec:ceiling}; then, in section \secref{sec:blueprint}, we propose a framework that envisions the potential for deeply domain-oriented efforts to be a crucial site for participation in the foundation model ecosystem.

\subsection{The participatory ceiling}
\label{sec:ceiling}
Looking across the approaches described above, we find the majority exhibit consistent limitations in the mode of participation they achieve. 
Why are participatory approaches thus far limited to those that lend little power to stakeholders?  
We argue that the intersection of primarily corporate control and context-agnostic models leads to a \textit{participatory ceiling} on what is possible when attempting to directly intervene on a foundation model.


\subsubsection{Foundation model developers currently lack incentives to share control with communities}
\label{sec:incentives}
In our analysis, the stakes of participation are most often limited to \textit{consultation}.
Participatory engagements primarily produce knowledge for model adjustments, rather than informing higher-level decisions around data sources or system purpose. 
Contributing to a ruleset or a red team does not guarantee one's input winds up reshaping the final model (\secref{sec:rulesets}), and contributing to a domain-specific evaluation does not guarantee others in one's community will not be harmed by products the model is based on (\secref{sec:domainharms}).



We argue this lack of capacity to govern models among communities contributing to them occurs, in part, because the current ecosystem of foundation models is massively centralized, resting primarily in the hands of well-resourced technology companies that amass the data and compute to deploy models.  
In any setting, participatory processes rarely guarantee outcomes---e.g., in democratic voting, participants usually do not decide how and when to vote, and are not guaranteed that their position will emerge as the majority. But the primacy of \textit{proprietary} models managed by corporate actors creates an additional layer of detachment separating public stakeholders from decision-making processes.  
If meaningful participation requires the distribution of \textit{some} decision-making power, this shift is not easily managed by firms, which are primarily constituted to protect shareholder interests rather than open collaboration to societal stakeholders. 
In recent history, large technology firms have especially emphasized they are not liable for downstream individual uses of their products;
when a firm does not have legal liability for harm, it has even less incentive to collaborate with impacted communities to identify and mitigate risks that may arise once technologies enter everyday lives. 
Moreover, most large organizations are risk-averse: partnerships with advocacy groups or community  advisory boards create openings for the company to be publicly criticized, or exposed to federal regulation and enforcement. 
There are also practical challenges: contending with intellectual property claims and integrating diverse user feedback into engineering are both known challenges across responsible AI work \cite{groves2023going}.

More broadly, the methods we surveyed 
tend to reflect an approach to "participation" that views model builders as the arbiters of representativeness, responsible for eliciting preferences from people and shaping the model \textit{for} them. 
In CCAI, for example, Anthropic developers culled its list of crowdsourced constitutional principles by selecting only those which displayed high public agreement, and eliminating those which were more controversial (e.g., whether the model should prioritize individual or collective good) \cite{anthropic2023ccai}.
This is not to say that developers do not have expertise that should be respected; the issue is that they remain in control of deciding what counts as `foundational' in a foundation model.  

Participants' lack of meaningful governance is exacerbated by corporate control of foundation models---but corporate dominance is not the sole issue here.
As we argue in the next subsection and unpack further in \secref{sec:discussion-power}, the fundamental premise of a foundation model approach assumes the need for \textit{a} centralized entity---be it a corporation, an academic institution, a government, or an open-source community---to orchestrate model development.
This core assumption separates model building from the contexts it aims to reflect, producing a disjointed supply chain of datasets in its wake.






\subsubsection{Meaningful participation necessitates context-specificity, but foundation models aim for universality}
\label{sec:context}
Because foundation models are intended to be applicable across domains, geographies, and other contexts, most of the participatory efforts we found aimed to construct general and universally applicable guidelines. The participatory tradition, however, has historically been grounded in \textit{context-specificity}. Prioritizing context in fields such as Participatory Action Research (PAR) stems from the acknowledgement that local stakeholders (i.e., people embedded in the day-to-day of a particular context) hold complex and valuable expertise about the needs, dynamics, and intricacies of their environment \cite{greenwood2006introduction}.  It also acknowledges that local knowledge systems are differentiated; values or norms in one context may not neatly transfer to another.  The same holds for understanding how harm manifests --- as described in intersectional feminist theory, the reality of marginalization is highly varied \cite{collective1977combahee, collins2019intersectionality, crenshaw2013mapping,d2023data}.  Considering context lends concreteness to \textit{what} harm actually looks like in a grounded domain and \textit{who} may be disproportionately affected.  
Finally, from the perspective of human-AI interaction design, deliberations around abstract or unbounded capabilities pose unique challenges for user-centered design processes \cite{yang2020re}. 
It becomes difficult for potential users to anticipate and reason about system behavior: e.g., empirical work has shown that the ways people react to and think about AI systems changes depending on whether they are engaged in an abstracted proxy task versus a more realistic, application-grounded task \cite{buccinca2020proxy,doshi2018considerations}.




When participatory efforts aim to produce universal outputs, they no longer prioritize local knowledge.  Instead, they ask participants to imagine and then reason about hypothetical, distant, or abstract scenarios. Consider a high-school teacher prompted to reason about whether ``being respectful'' or ``conveying clear intentions'' is a more important value for an LLM to adhere to (both examples from CCAI \cite{anthropic2023ccai}), versus considering the concrete tradeoffs and risks of an LLM-based tutoring tool integrated into their classroom. 
While participants may be perfectly capable of abstraction and imagination, foundation models are subject to tremendous hype under the banner of artificial general intelligence, and their governance requires a willingness to speculate on risks that may be quite far from participants' lived realities---a unique challenge within human-AI interaction \cite{yang2020re}. 
Such speculation might lead to greater technological awareness and literacy, but as \citet{harrington2019deconstructing} argue, asking marginalized communities to engage in the ``\textit{blue-sky ideation}'' of technology design risks ultimately frustrating underserved individuals.
By focusing instead on the real harms and concerns people are presently experiencing, developers can limit the material and affective demands of participation \cite{dourish2020being}, acknowledge participants as experts on local knowledge systems, and create outputs that more closely reflect actual downstream needs. 

\paragraph{Pushing the participatory ceiling}
Notably, in our review of the current state of play, a few examples out of those surveyed stood out in straining against this ceiling. For example, UbuntuAI \cite{eglash2023ubuntu, eglash2024computational} proposes a system that responds to the expropriation by foundation models of African artists' intellectual property by collaboratively creating a licenseable dataset of their work.  While the project is still exploratory, its approach touches on the \textit{collaboration} mode of participation, with ongoing co-creation of the dataset and a compensation structure that lends itself to community control.
Here, gains on the parameters of participation are realized, in part, through a focus on a specific domain. In the next section, we draw out the implications of examples like this into a broader framework for more meaningful participation into foundation models.

\section{A Blueprint for More Participatory Foundation Models}
\label{sec:blueprint}

We propose a three-layer framework to enable more effective public participation in foundation models.  
As discussed in the prior section, we claim there is a fundamental tension between the scale and generalizability of foundation models, and the power that local communities can wield in shaping them.  Our framework addresses this limitation by building in additional layers for participation at more local, application-oriented scales---the \textit{subfloor} layer and the \textit{surface} layer.  The naming of our framework builds on the metaphor of a \textit{foundation}: an unfinished base upon which many different kinds of structures might be built.  The \textit{subfloor} is a stable and level ground built on the foundation, which provides a structurally sound base for any number of top-level \textit{surfaces}.  


In this section, we first describe the three-level blueprint.  We then delve into three case studies, illustrating opportunities for participation it affords when applied to foundation model usage in healthcare, banking, and journalism. Each case study, while hypothetical, is grounded in current uses of LLMs, real-world organizations, and discussions with a domain expert in each context.  We intend the case studies to serve as starting points for further iteration and testing out in practice. 

\subsection{The Blueprint}

\begin{figure}
    \includegraphics[width=\columnwidth]{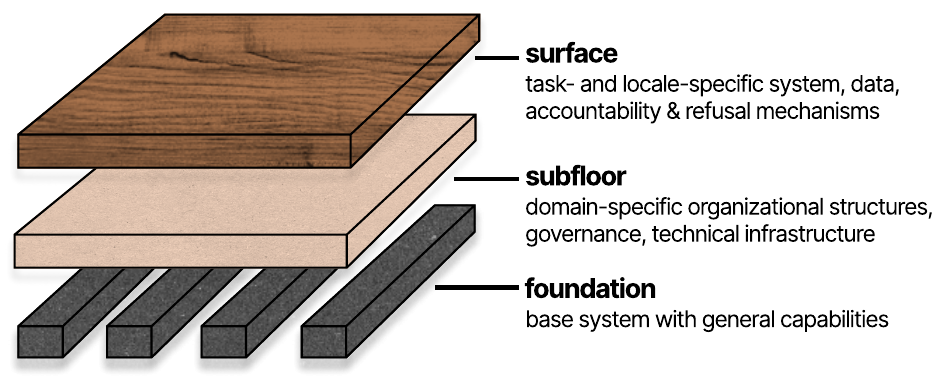}
    \caption{An illustration of the three layers proposed in our blueprint.}
    \label{fig:enter-label}
\end{figure}

\subsubsection{The Foundation Layer}
The foundation layer is where a base model 
---not yet intended to be domain- or application-specific---is created and maintained.   
Our analysis of current attempts at participation in foundation models (section \secref{sec:ceiling-overall}) shows
there are already numerous approaches used and stakeholders involved; 
still, there are certainly avenues to deepen this work with respect to participatory desiderata.
For example, 
foundation model developers might pursue forms of participation that are ongoing and sustainable, 
or widen the scope of interventions to those that impact other parts of the lifecycle beyond the steering of the model.  

That said, we contend that participation at the foundation layer \textit{is not, by itself, enough}, due to inherent limitations in what is possible at scale (\secref{sec:ceiling}).
The subfloor and surface layers of our framework are instead grounded in context, and necessary for tractable participation.  
In this setup, the foundation layer takes on additional responsibilities: ensuring the base functionality required by the network of subfloors and surfaces, and remaining porous to issues or demands raised by them (e.g., compensating and attributing creative workers for the use of their work, as in \secref{sec:case-journo}).




\subsubsection{The Subfloor Layer}
The subfloor layer encompasses technical infrastructure, norms and governance for a grounded domain. This might include fine-tuned models, curated datasets, auditing process, mechanisms for recourse, standards-setting procedures, and shared governance structures (we expand on specific instantiations of the subfloor layer in Sections \ref{sec:case-health}-\ref{sec:case-finance}).  Importantly, participation at the subfloor 
mitigates the pitfalls of generality (\secref{sec:context}) by being grounded in a specific domain, and the barriers of corporate incentives (\secref{sec:incentives}) via ownership by entities such as nonprofits, local governments, community advocacy organizations, or unions.  Context-specificity makes more clear \textit{who} should be involved: for instance, in the healthcare domain, participatory efforts might engage medical practitioners, health equity scholars, health policy makers, patient advocacy groups, and voices from communities we know to be marginalized within the current healthcare system \cite{pierson2023use}.  It also becomes easier for people and communities to participate.  Rather than coming up with abstract, universal desiderata (``the model should produce outputs that are good for humanity''), the scope of harms to consider is bounded by a particular domain.  
Stakeholders can thus contribute their concrete expertise and lived experience of the domain (``the model should not replicate known biases in pain assessment'') \citep[e.g.,][]{pfohl2024toolbox, zack2024assessing}.    


\subsubsection{The Surface Layer}
The surface layer corresponds to a specific downstream use case, built on top of the subfloor layer.  This might encompass a specific tool or system, along with task-specific datasets(s), documentation, and mechanisms for accountability and refusal.  
Participation at the surface layer likely looks similar to existing case studies of participatory ML.
Affected communities can shape problem formulation, and co-determine whether a foundation model-based solution is desirable in their context.  For example, a specific healthcare facility might want a tool to share information to patients on reproductive health. 
Local patients and care providers, along with reproductive justice experts, patient advocacy groups, and other stakeholders, might collectively decide on the appropriateness of an LLM-based tool. 
If the outcome of this deliberation is to move forward, it might involve participatory data collection and annotation that is specific to the task and locality; this data can be used for fine-tuning the subfloor model or for context-specific validation.  
Participation at this layer might further involve avenues for public accountability and recourse, determined and designed by the affected stakeholders. 

Importantly, the existence of the subfloor layer means that the burden of domain-specific validation and specialization is not put entirely on stakeholders developing a particular use case. 
For example, our local healthcare facility can inherit the baseline equity assurances put in at the subfloor layer (``the model should not replicate known biases in pain assessment''), while building specialized constraints for their patient population (``the model should accurately answer questions that are most common among our patient population'').  We elaborate on a related example in \secref{sec:case-health}.

\subsection{Case study 1: Clinical care}
\label{sec:case-health}
Our first case study illustrates the participatory opportunities provided by the nested subfloor-surface structure of our framework.  We consider the development of LLM-based tools that aim to address growing administrative burdens clinicians face due to electronic health records (EHRs) \cite{gawande2018doctors}.  Such tools may transcribe the doctor-patient interaction with a speech-to-text model; these transcriptions may then be used for clinical note summarization.  This paradigm (also called ``\textit{ambient intelligence}'') is increasingly used in dominant EHRs like Epic \cite{epic2023generative}.

\paragraph{The problem} 
A tool for transcribing medical interactions must robustly handle a diversity of languages, accents, and idioms, without inadvertently reproducing harmful stereotypes about specific minoritized groups, or compromising medical accuracy. There are also issues of trust and privacy---patients and providers must be able to verify a summary before it becomes part of the record, and to opt out if they do not want the tool listening in the exam room.


Participatory approaches could improve these systems with respect to equity and trust.  
However, these concerns are hard to address at the foundation layer: they require engaging with complex and domain-specific issues at the intersection of dialect, geography, race and ethnicity, service provision, and medical history. 
Still, if the responsibility of thoroughly auditing a model were to rest with individual clinics, they would each
likely run into data limitations (e.g., for under-represented dialects), and may needlessly duplicate effort 
across different settings.

\paragraph{The opportunity and participatory methods} 
Here, the subfloor layer could encompass a constellation of patient advocacy organizations who pool data collected with communities to represent various dialects and accents, as well as identify key risks to anticipate and audit.
For example, a patient advocacy organization in a predominantly Black community might collect data representing African American Language (AAL) \cite{deas2023evaluation}. 
They might further identify Black maternal health as a key equity risk, and elevate the perspectives of Black midwives to shape the data collection process and identify specific harms to audit.  Similarly, an organization representing an Asian-American immigrant community might collect data representing the different kinds of accents and idioms in that community, and flag Asian-American mental health as an important equity risk.

Data collected across communities could be used to fine-tune a shared speech-to-text model 
that performs well on speech from both of these communities, and to design ongoing audits of this model for both sets of identified risks.
The pooled data could include data and deliberation from as many communities as is relevant to the subfloor's identified constituency.
Additional expertise (e.g., on legal, privacy, or health equity issues) could also shape technical infrastructure and broader guidelines for use.
Such a convening might build, for example, on the format proposed in \citet{antoniak2023designing}, where healthcare workers and birthing people collaborated on guiding principles for the use of NLP in maternal health.



At the surface layer, then, resources made available at the subfloor (e.g., guidelines for use, audits across equity dimensions) can inform concrete deliberation at a specific care site around whether an LLM-based transcription tool is desired in that context. The surface layer thus provides an important site of refusal, for the opt-out and oversight mechanisms important to ensure trust.  If the outcome of deliberation is to move forward, the surface layer provides an opportunity to incorporate context-specific logic and validation around, e.g., the resources, procedures, and patient populations at a given hospital, while building on the technical infrastructure and trust established at the subfloor.

\subsection{Case Study 2: Journalism}
\label{sec:case-journo}
In this case study, we examine opportunities for collective stakeholder power at the subfloor to shape changes at the foundation layer.  We consider this opportunity within the context of foundation models' impact on professional writers and other creatives. We focus on those employed by newsrooms---organizations which themselves assert a copyright claim over their material.

\paragraph{The problem} In January 2024, the New York Times filed a lawsuit against OpenAI and Microsoft for the unauthorized use of the publication's copyrighted material to train AI systems \cite{grynbaum2023times}. 
Several other organizations and individuals have filed similar complaints objecting to the wide-scale collection and use of copyrighted material without compensation or credit \cite{alter2023franzen,small2023sarah}. The foundation model-based systems in question further risk leading to labor displacement of the same people whose work was expropriated to create it. 
When trying to enact changes in the data used at the foundation layer (for example, by filing individual lawsuits), each individual creator lacks enough leverage to shift data collection norms. The prohibitive cost of challenging this practice in effect establishes a precedent that data misuse is permissible. 

\paragraph{The opportunity and participatory methods} 
The subfloor layer presents an opportunity for entities with copyright concerns to partner with each other, collectively asserting copyright claims over their data and establishing a license agreement for compensation over time. Such an effort might be housed within an existing collective action organization, such as the NewsGuild, a primary union for media workers. Inspired by \citet{eglash2023ubuntu}, the subfloor could support the creation of a curated dataset of license-ready contributions, alongside organizational infrastructure that would keep it updated with successively published material. Collectively, publishers and creators could demand the expungement or filtering of their data from existing datasets used to train foundation models. They could then establish an ongoing data use agreement and payment structure such that the data cannot be used without compensation. Such a setup might be supported by technical approaches such as SILO \cite{min2023silo}, which could enable creators to opt in/out of a datastore used only during inference and be attributed for the use of their work.  In this case, the subfloor leverages collective action to influence changes at the foundation layer that reflect domain-specific needs and requirements. In its absence, individual lawsuits may both duplicate efforts and lack the power needed to demand a compensation agreement.

\subsection{Case Study 3: Financial services}
\label{sec:case-finance}
Finally, we consider how the subfloor layer could provide avenues for participation in contexts such as finance, where major financial firms are exploring the use of LLMs for improving fraud risk scoring algorithms \cite{mastercard, visa, Lepeska}. While machine learning techniques are already used for assigning risk scores to transactions, LLMs are beginning to be used to identify more subtle patterns in fraudulent transactions, and may introduce new and unforeseen risks.

\paragraph{The problem} As a task, fraud detection carries high-stakes fairness and equity risks. To avoid reputational damage, credit card issuing banks and payment networks aim to avoid declining transactions frivolously or inequitably. They are bound by legal requirements to protect equity as well: in the U.S., the Federal Trade Commission and the Equal Credit Opportunity Act hold credit card companies accountable for credit discrimination. 

Robust reporting mechanisms can help to catch new risks and unfair patterns of credit declination in deployment; however, competitiveness concerns incline firms to conduct system evaluation strictly behind closed doors. As a result, potential harms like false positives in fraud detection and inequitable credit service provision remain under-examined---particularly by the communities who would be most affected and the organizations that represent them. 
Because these concerns are deeply domain-specific, addressing them at the foundation layer proves difficult.
If ensuring equity is left to individual providers, however, we risk leaving affected communities to navigate consequences downstream of a complex chain of systems that each implement fraud detection technology differently.


\paragraph{The opportunity and participatory methods} 
The subfloor in this case could focus on robust on-the-ground reporting and recourse mechanisms to catch new or unexpected patterns of discrimination. For example, participatory reporting mechanisms might include: (1) \textit{in vivo} field testing ``spot-checks'' of the payment system with end-users from high risk groups; (2) reporting hotlines by which the broader public can escalate their experiences and concerns; and (3) formal reports of concerns and common experiences from various constituencies. 

Working groups of experiential experts and advocates can then use this input to inform foundation model governance in financial services---identifying larger patterns in model behavior, documenting harms and vulnerabilities, and producing updated guidance.  These working groups can be organized around an institutional body already common to financial services providers, such as the Payment Card Industry (PCI) Security Standards Council, which convenes and issues guidance to financial services firms on information security and privacy protections. Its remit is also inclusive of best practices for new tools and technologies, such as generative AI. Guidance produced by working groups can inform PCI member firms' in-house counsel, privacy and risk management personnel, and regulatory teams. These audiences can then translate findings for their own firm's product teams. 

Together, these reporting and organizational infrastructures would create avenues for more robust, real-world, and variegated external input to shape surface-layer systems, in spite of a highly closed and rigid overall environment.
It would also serve as an ``early warning system'' when cardholder communities or merchants face an uptick in declined transactions because of spurious or discriminatory correlations made by an LLM-backed product.



\section{Discussion}
We have presented a conceptual contribution for how meaningful participation can shape the foundation model lifecycle.  Our blueprint supports shifting power within the foundation model ecosystem through a focus on context. 
We hope our work helps organize the FAccT community's thinking around this new and rapidly growing paradigm.
But participation is not meant as a panacea, and our framework inherits its familiar risks and limitations: diffusion of accountability (\secref{sec:discussion-accountability}), power asymmetry and co-optation (\secref{sec:discussion-power}), and labor for already disempowered stakeholders (\secref{sec:discussion-refusal}). 
In this section, we unpack each of these limitations, and outline future work for the FAccT community addressing these risks.

\subsection{Accountability through the subfloor layer}
\label{sec:discussion-accountability}

Our case studies highlight a wide range of participatory mechanisms afforded by the subfloor, including collaborative data collection and model auditing (\secref{sec:case-health}), collective action for data ownership and refusal (\secref{sec:case-journo}), and avenues for on-the-ground reporting and recourse (\secref{sec:case-finance}).
In each case, the subfloor provides a route for meaningful participation by creating more tractable accountability relationships between the foundation model provider and downstream surface layers.
Rather than engagement that is atomized and discrete (as we see with efforts at the foundation layer), because it is much closer to---if not directly controlled by---affected communities, the subfloor enables participation that is \textit{deliberative} and \textit{longitudinal}.  
It breaks the participatory ceiling through being grounded in context and primarily owned by entities such nonprofits, local governments, community advocacy organizations, or unions. 
Our framework shifts the locus of responsibility from a central model creator to an ecosystem of subfloors, enabling a domain-specific actor to become accountable for a system. At the same time, collective action by subfloors, who together can represent a substantial stake in an industry, 
becomes harder for a foundation layer to brush off.  The impact of this engagement scales across surface-layer use cases, which provide sites for additional, task-specific participatory processes while inheriting trust built at the subfloor. 

Still, a layered framework diffuses accountability for harm in ways that scholars have yet to resolve.
The relationship between the foundation layer and the subfloor layer may resemble that of an app store provider to its app developers. 
Apple's Terms of Service define guidelines for apps to satisfy to be hosted in their marketplace; a failure to meet these requirements results in removal from its App Store, and thus a limited channel for distribution to iOS users. 
But how does a foundation model become responsible (or not) for harms it enables at the subfloor and surface layers? 
And what responsibilities do subfloors have for harms they may inherit from foundations, or host and pass on to surfaces on their own?
More participatory and community-controlled infrastructure may enable bad actors to fine-tune foundation models for abusive ends
\cite{oremus2023elon}.
The stakes of this debate are rising: as a stark example, it remains unclear who ought to be responsible for an LLM-powered chatbot encouraging a person to harm themselves \cite{xiang2023death}.
Understanding where liability for harm can and should rest will be key to enabling participatory infrastructure while mitigating its drawbacks. 

\subsection{Centralization and the limits of transparency}
\label{sec:discussion-power}

Our analysis identified the inherent centralization in foundation model development as a key bulwark of the participatory ceiling (\secref{sec:ceiling}).
Whether a venture-backed startup, a technology giant, or a well-resourced academic or government institution, foundation model developers are an obligatory passage point for influencing the operation of a given foundation model. 
Crucially, this is a feature of the foundation model paradigm that remains consistent across open-source and closed-source approaches.
Many researchers have advocated for transparency and openness in foundation models: whether model weights and data are available for inspection, or whether a corporate or nonprofit entity hosts a given model \cite{bommasani2023transparency}. 
We contend, however, the issue is more in the fundamental premise that models can be disconnected from meaningful social governance, even as they aim to represent the complexity of language, moral reasoning, and other human social interactions. 
While the precedents around radical transparency set by free and open-source software communities are an important first step towards meaningful participation in \textit{development}, 
Widder et al. \cite{widder2023open} have described how in practice, the sheer scale of resources needed to \textit{deploy} a foundation model mean only a handful of well-resourced institutions are positioned to engage.

It remains to be seen to what extent this inherent power asymmetry can be allayed by a subfloor. A subfloor with a large remit (e.g., to represent the English language) could recapitulate this centralization of power. 
Moreover, like all participatory mechanisms, the subfloor risks participation-washing by powerful actors in service of their own aims. \citet{ahmed2020we} warns that participation can be a means not of advancing collective well-being, but of powerful actors creating a heavily circumscribed mechanism for co-optation that impedes real change. 
At best, our blueprint raises the stakes for participation-washing at scale, and also creates a more effective target for advocacy and organizing.

To realize this best-case, we see ample future work understanding how to scope a subfloor or a surface so it remains accountable to communities' needs.
In our clinical care case study (\secref{sec:case-health}), for example, a subfloor must take on the responsibility of coordinating across different communities, each with their own goals and capacities, to ensure equitable performance and risk mitigation in a common speech-to-text model.
What constitutes a subfloor's domain, and how should the boundaries between subfloors be determined? 
How can we collectively place constituencies between subfloors, and ensure common deliberative processes among them?
Technical advancements in securely pooling data and training shared models are also important, and could build on existing approaches like federated learning and open science.\footnote{e.g., in clinical care, Nightingale (https://www.ngsci.org/), Observational Health Data Sciences and Informatics (OHDSI) (https://www.ohdsi.org/), and the Patient-Led Research Collaborative (https://patientresearchcovid19.com/).} 
Progress on these fronts will help create the healthy ecosystem of subfloors and surfaces we need to establish decentralization and accountability in foundation models' future.

\subsection{Ownership, refusal, and the burden of participation}
\label{sec:discussion-refusal}

Finally, there is the issue of how to make meaningful participation manageable for stakeholders---especially those who may already face marginalization and disempowerment.
Better opportunities for participation still require time and energy of people who may not want to donate their time to governance efforts; particularly since so much of foundation model governance involves the labor of collecting and managing large-scale datasets. 
For participation in foundation models to achieve lofty aims of redistributing power to the marginalized, we need further work updating mechanisms for data stewardship and consent for the foundation model era.

These mechanisms are well-suited to subfloor and surface layer interventions. We are encouraged to see growing interest in how individuals and communities can exercise agency over the data collection and modeling processes underlying machine learning, via principles ranging from data refusal \cite{zong2024data} and participatory data stewardship \cite{tseng2024data, lovelace2021participatory} to AI contestability \cite{denton2020bringing, kluttz2022shaping}.
The need for individuals to have control over the data that flow through foundation model-based systems has also begun to motivate new model architectures: e.g., techniques like SILO are explicitly motivated by the need to isolate sensitive data from a model, so it is used in inference but not training, and so individuals have the agency to remove their data from what the model can use \cite{min2023silo}.
In our framework, we envision that surface layer interventions could offer a space for individual data refusal; and subfloor layer interventions could offer a space for collective contestation around data usage. 

Importantly, we hope that our framework leads to more participatory foundation models---but also lends communities greater agency in refusing the use foundation models at all.  We envision an important part of participation at the surface layer, for example, involving deliberations about if a foundation model is the right or desired approach to a given problem.  Domain-specific guidelines or audits at the subfloor layer can help ground this deliberation, but do not guarantee that the outcome will be to move forward. In other words, the blueprint is not an endorsement or guarantee of foundation model usage.  Rather, it creates more participatory alternatives, but holds that control around downstream usage should always rest with those closest to a specific context. 

\section{Conclusion}
In this paper, we examined the intersection between participatory approaches and foundation models. Because foundation models are developed for use across a wide array of settings, any individual group seeking to provide input into the foundation model or mitigate harms on-the-ground must vie for influence amid a vast number of similarly impacted parties for the time and resources of a single developer firm. This convergence of scale and power asymmetry exacerbates the challenges of shaping foundation models with public input. To better support meaningful participation over foundation models, we define an organizational-level intervention that could address this challenge: a “subfloor” layer supports more local, application-oriented opportunities for meaningful participation in the form of shared technical infrastructure, norms, and governance for a grounded domain such as journalism or financial services. The subfloor layer scopes the range of potential harms to consider and affords individual participants more concrete avenues for intervention, scaling that engagement to downstream surface layer use cases.

\begin{acks}
We would like to thank our colleagues, peer reviewers, and subject matter experts who have been so generous with their feedback and insight. In particular, we are grateful to Robert Lewis, Seth Lazar, Jon Kleinberg, Mark Sendak, Merill Fabry, Albert Gehami, the students and faculty of Australia National University's School of Cybernetics, and presenters at the Stanford Sociotechnical AI Safety workshop. This work is also supported in part by the John D. and Catherine T. MacArthur Foundation, a Microsoft Research PhD Fellowship (ET), and the Principles G.I. Coffee House in Gowanus, Brooklyn.

\end{acks}

\bibliographystyle{ACM-Reference-Format}
\bibliography{references}

\appendix

\end{document}